\documentstyle [12pt]{article}
\makeatletter \oddsidemargin  0in \evensidemargin 0in
\textwidth16cm \RequirePackage[dvips]{graphicx} \textheight 20.5cm
\newcommand{\singlespacing}{\let\CS=\@currsize\renewcommand{\baselinestretch}{1.5}\tiny\CS}
\makeatletter \oddsidemargin  -.1in \evensidemargin -.1in
\newcommand{\doublespacing}{\let\CS=\@currsize\renewcommand{\baselinestretch}{1.35}\tiny\CS}

\def\@citex[#1]#2{\if@filesw\immediate\write\@auxout{\string\citation{#2}}\fi
  \def\@citea{}\@cite{\@for\@citeb:=#2\do
    {\@citea\def\@citea{,\linebreak[0]\hskip0pt plus .2em}%
      \@ifundefined{b@\@citeb}%
    {{\bf ?}\@warning{Citation `\@citeb' on page \thepage\space undefined}}%
      \hbox{\csname b@\@citeb\endcsname}}}{#1}}

\newtheorem{rule-def}[theorem]{Rule}
\begin{document}
\title{Impossibility of Probabilistic splitting of quantum information}
\author{I.Chakrabarty $^1$ $^2$,\thanks{E-mail: indranilc@indiainfo.com}, B.S.Choudhury $^2$\\ $^1$ Heritage Institute of Technology, Kolkata,
India\\ $^2$ Bengal Engineering and Science University, Howrah,
India}
\date{}
\maketitle{}
\begin{abstract}
We know that we cannot split the information encoded in two
non-orthogonal qubits into complementary parts deterministically.
Here we show that each of the copies of the state randomly
selected from a set of non orthogonal linearly independent states,
splitting of quantum information  can not be done even
probabilistically. Here in this work we also show that under
certain restricted conditions, we can probabilistically split the
quantum information encoded in a qubit.
\end{abstract}
\section{Introduction :} In quantum information theory
understanding the limits of fidelity of different operations has
become an important area of research. Noticing these kind of
operations which are feasible in classical world but have a much
restricted domain in quantum information theory started with the
famous 'no-cloning' theorem [1]. The theorem states that one
cannot make a perfect replica of a single quantum state. Later it
was proved that one cannot clone two non-orthogonal quantum
states [2]. However this does not rule out the possibility of
producing approximate cloning machines [3-5]. Even though
deterministic cloning is not possible, a probabilistic cloning
machine can be designed , which will clone the input states with
certain probabilities of success [6]. Though the unitarity of
quantum mechanics prohibits accurate cloning of non orthogonal
quantum states , but such a class of states secretly chosen from
a set containing them can be faithfully cloned with certain
probabilities if and only if they are linearly independent.
 Basically Quantum copying machine can be divided into two classes
(i)deterministic quantum copying machine (ii) probabilistic
quantum copying machine. The first type of quantum copying
machine can be divided into two further subclasses: (i) State
dependent quantum cloning machine , for example,  Wooters-Zurek
(W-Z) quantum cloning machine [1], (ii) Universal quantum copying
machine, for example, Buzek-Hillery (B-H) quantum cloning machine
[2].\\
 Pati and Braunstein
introduced a new concept of deletion of an arbitrary quantum state
and shown that an arbitrary quantum states cannot be deleted. This
is due to the linearity property of quantum mechanics. Quantum
deletion [7,8] is more like reversible 'uncopying' of an unknown
quantum state. The corresponding no-deleting principle does not
prohibit us from constructing the approximate deleting machine
[16]. J. Feng et.al. [9] showed that each of two copies of
non-orthogonal and linearly independent quantum states can be
probabilistically deleted by a general unitary-reduction
operation. Like universal quantum cloning machine, D'Qiu [15] also
constructed a universal deletion machine but unfortunately the
machine was found to be non-optimal in the sense of fidelity. A
universal deterministic quantum deletion machine is designed in an
unconventional way that improves the fidelity of deletion from 0.5
and takes it to 0.75 in the limiting sense [19]. Many other
impossible operations generally referred as
 'General Impossible operations' [10] can not be achieved successfully
 with certainty , but one can carry out these operations at least
 probabilistically with certain probability of success  [11].\\
 Recently 'no-splitting' theorem [12] is an
 addition to these set of no-go theorems. The theorem states that the
 quantum information of a qubit cannot be split into
 complementary parts. The no-splitting theorem, can be mathematically
  stated as whether the two real parameters ($\theta,\phi$) can be split into two complementary qubits
  as follows: $L(|A(\theta,\phi)\rangle |B\rangle=|A(\theta)\rangle
  |B(\phi)\rangle)$? The answer is no. The linearity of quantum
  mechanics [12] as well as the unitarity of quantum mechanics
  doesn't allow the splitting of the information contained inside
  a qubit.

  Similarly 'partial erasure of quantum information'
 [13] is another operation which is not possible in quantum world.
 The 'no-splitting' theorem can also be obtained as a special case
 of 'no-partial erasure' of quantum information theorem. It remains interesting to see that
 whether we can split quantum information at least
 probabilistically. In this work we try to find out whether we can
 split the information in a qubit with a certain probability of
 success. We show that we cannot even probabilistically split the
 quantum information inside a qubit. Here we will find that unlike cloning and deletion
 if non orthogonal quantum states
 are secretly chosen from a set then there exists no such
 transformation that will split the quantum information of a qubit
 with certain probability of success. However we also show that
 under certain restricted conditions the splitting of quantum
 information will be possible.

 \section{Probabilistic Quantum information splitting:} The quantum
'no-splitting' theorem  [12] says that exact splitting of quantum
information encoded in two non orthogonal states cannot be done.
Nevertheless, it does not get rid of the possibility of splitting
the quantum states with certain probabilities or in other words
one may ask that is there any unitary reduction process which will
split the information encoded in two non-orthogonal states
 $|\psi_i (\theta_i,\phi_i)\rangle$ and $|\psi_j (\theta_j,\phi_j)\rangle$
secretly chosen from a set   $ S=\{
|\psi_1(\theta_1,\phi_1)\rangle , |\psi_2
(\theta_2,\phi_2)\rangle,.... |\psi_n
(\theta_n,\phi_n)\rangle$\}. Here $|\psi_i
(\theta_i,\phi_i)\rangle=\cos(\frac{\theta_i}{2})|0\rangle+
\sin(\frac{\theta_i}{2})e^{\imath\phi_{i}}|1\rangle$ are the
quantum states represented as a point on the Bloch sphere (where
$\imath=\sqrt{-1}$ ). Let us consider two systems A and B . Now
each of the states of the set S can be taken as a input state of
the system A. Let us consider a unitary evolution U and
measurement M, which together yield the following evolution
\begin{eqnarray}
|\psi_i (\theta_i,\phi_i)\rangle|\Sigma\rangle
\longrightarrow^{[U+M]}\longrightarrow |\psi_i
(\theta_i)\rangle|\Sigma (\phi_i)\rangle
\end{eqnarray}
where  $ |\Sigma\rangle $ is the input state of the ancillary
system B. Both the systems A and B are described by a N
dimensional Hilbert space with $ N\geq n $ .\\

To continue with the argument of the above statement, a probe P
with $n_p$ ( $n_p\geq n+1$ ) dimensional Hilbert space is
introduced , where $ \{ |P_0\rangle, |P_0\rangle, .....,
|P_n\rangle \} $ are n+1 orthonormal states of the probe. Now let
us introduce a unitary operator U whose action on the tensor
products of the Hilbert spaces associated with the system A,B and
probe P is given by
\begin{eqnarray}
U(|\psi_i
(\theta_i,\phi_i)\rangle|\Sigma\rangle|P_0\rangle)&=&\sqrt{\gamma_i
}|\psi_i (\theta_i)\rangle|\Sigma (\phi_i)\rangle|P_0\rangle
{}\nonumber\\&&+\sum_{j=1}^{n}
c_{ij}|\Phi_{AB}^{(j)}\rangle|P_j\rangle,~~~(i=1,2,..,n)
\end{eqnarray}
where $|\psi_i
(\theta_i)\rangle=\cos(\frac{\theta_i}{2})|0\rangle+\sin(\frac{\theta_i}{2})|1\rangle$
and $|\Sigma
(\phi_i)\rangle=\frac{1}{\sqrt{2}}[|0\rangle+e^{\imath\phi_i}|1\rangle]$.
Here, $\{|\Phi_{AB}^{(j)}\rangle\}$ (j=1,...,n) are normalized
states of the composite system AB, and these states are not
necessarily orthogonal. After the unitary evolution , the
measurement is made on the probe P . The attempt made for
splitting the information into constituent parts will succeed with
$\gamma_i$ probability of success if the measurement outcome of
the probe is
$P_0$.\\
We start here by showing that unlike probabilistic cloning and
deletion , probabilistic splitting of quantum information will
not be possible  for both linearly dependent and independent
states secretly chosen from the set S.
Let us introduce a theorem.\\\\
\textit{Theorem1 } : \textbf{The states which are secretly chosen
from the set $S=\{|\psi_1(\theta_1,\phi_1)\rangle ,\\ |\psi_2
(\theta_2,\phi_2)\rangle,...|\psi_n (\theta_n,\phi_n)\rangle\}$
can be probabilistically split (realization of the unitary
evolution)  if the states are linearly independent}
.\\
\textit{Proof} :  Consider an arbitrary state which can be
expressed as  the linear combination of the states in the set S.
\begin{eqnarray}
|\psi(\theta,\phi)\rangle=\sum_{i=1}^{n}d_{i}|\psi_i(\theta_i,\phi_i)\rangle
\end{eqnarray}
The unitary transformations of the arbitrary linearly dependent
state vector is given by,
\begin{eqnarray}
U(|\psi(\theta,\phi)\rangle|\Sigma\rangle|P_0\rangle)=\sqrt{p}|\psi(\theta)\rangle|\Sigma(\phi)\rangle
|P_0\rangle +c|\Phi_{AB}\rangle|P_{\perp 0}\rangle
\end{eqnarray}
But, if we consider the action of the unitary transformation
defined in (2) on the linear combination of the state vectors
belonging to the set $S$, then the resultant is given by,
\begin{eqnarray}
U(\sum_{i=1}^{n}|\psi_i(\theta_i,\phi_i)\rangle
|\Sigma\rangle|P_0\rangle )=
\sum_{i=1}^{n}\sqrt{p_{i}}d_{i}|\psi_i(\theta_i)\rangle
|\Sigma(\phi_i)\rangle|P_0\rangle + \sum_{i=1}^{n}\sum_{j=1}^{n}
d_{i}c_{ij}|\Phi_{AB}^{(j)}\rangle|P_j\rangle
\end{eqnarray}
Now it is clearly evident that the final states (4) and (5) are
different quantum states . Since the state
$|\psi(\theta,\phi)\rangle$ is a linear combination of the state
vectors $|\psi_i(\theta_i,\phi_i)\rangle$ belonging to the set
$S$, the linearity of quantum mechanics is prohibiting the
existence of probabilistic quantum information splitting machine.
Therefore the unitary evolution given by (2) exists for any set
secretly chosen from the set $S$ only if the states belonging to
the set $S$ are linearly
independent.\\\\
The interesting part is that the converse of the theorem is not
true. The converse statement of the \textit{Theorem1} is given as
follows :\\
 \textbf{If the quantum states $|\psi_i(\theta_i,\phi_i)\rangle
~~~(i=1,..n)$ in the set $S$ are linearly independent then the
unitary evolution (2) exists}. \\

 However we will find that this
will not hold in general. Interestingly we will also see that
there are few particular cases for which the converse is true, and
consequently the information splitting is possible.
\\\
In other words we can say that if the states chosen secretly are
linearly independent, the unitary evolution (2) will not hold with
positive definite matrices $\sqrt{\Gamma}$, consequently the
physical process described by (1) is not going to be realized. In
order to verify the existence of the unitary evolution (2) we must
introduce the
following lemma.\\\\
\textit{Lemma1}: If two sets of states $\{|X_1\rangle,
|X_2\rangle,.....,|X_n\rangle\}$ and
$\{|\tilde{X_1}\rangle,|\tilde{X_2}\rangle,.....,|\tilde{X_n}\rangle\}$
satisfy the condition
\begin{eqnarray}
\langle X_i|X_j\rangle= \langle
\tilde{X_i}|\tilde{X_j}\rangle~~~~(i=1,...n;j=1,...n)
\end{eqnarray}
there exists a  unitary operator U to make
$U|X_i\rangle=|\tilde{X_i}\rangle~~~~(i=1,...,n)$\\\\
The $n\times n$ inter-inner products of equation (2) yield the
matrix equation
\begin{eqnarray}
 D=\sqrt{\Gamma}GH\sqrt{\Gamma}^{+} + CC^{+}
\end{eqnarray}
where $D=[\langle
\psi_i(\theta_i,\phi_i)|\psi_j(\theta_j,\phi_j)\rangle]$,
$G=[\langle \psi_i(\theta_i)|\psi_j(\theta_j)\rangle]$,
$H=[\langle \Sigma(\phi_i)|\Sigma(\phi_j)\rangle]$ and
$C=[c_{ij}]$. The diagonal efficiency matrix $\Gamma$ is defined
by $\Gamma=diag(\gamma_1,\gamma_2,...,\gamma_n)$, hence
$\sqrt{\Gamma}=\sqrt{\Gamma}^{+}
=diag(\sqrt{\gamma_1},\sqrt{\gamma_2},...,\sqrt{\gamma_n})$. Now
if \textit{lemma1} clearly shows that the equation (7) is
satisfied with a diagonal positive-definite matrix $\Gamma$ ,
then the unitary evolution (2)will hold, consequently the
physical process (1) can be realized in
physics.\\
To show that there is a diagonal positive definite matrix
$\Gamma$ to satisfy equation (7), first we need to show that the
matrix $D$ is positive-definite.
\\We introduce a Lemma to show that $D$ is
positive-definite.\\\\
\textit{Lemma2}: If n states
[$|\psi_1(\theta_1,\phi_1)\rangle,|\psi_2(\theta_2,\phi_2)\rangle,....|\psi_n(\theta_n,\phi_n)\rangle$]
are linearly independent, then the matrix $D=[\langle
\psi_i(\theta_i,\phi_i)|\psi_j(\theta_j,\phi_j)\rangle]$ is
positive
definite.\\

\textit{Proof}: For any arbitrary n-vector
$B=(b_1,b_2,....b_n)^{T}$, the quadratic form $B^{+}DB$ can be
expressed as
\begin{eqnarray}
B^{+}DB=\langle\Psi|\Psi\rangle=\||\Psi\rangle\|^{2}
\end{eqnarray}
where
\begin{eqnarray}
|\Psi\rangle= b_1 |\psi_1(\theta_1,\phi_1)\rangle+
b_2|\psi_2(\theta_2,\phi_2)\rangle+...+ b_n
|\psi_n(\theta_n,\phi_n)\rangle
\end{eqnarray}
Since we know that states
[$|\psi_1(\theta_1,\phi_1)\rangle,|\psi_2(\theta_2,\phi_2)\rangle,....|\psi_n(\theta_n,\phi_n)\rangle$]
are linearly independent, the  state $|\Psi\rangle$ does not
reduce to zero for any n-vector B and its norm will always remain
positive. Hence from definition $D$ is
positive-definite.\\\\
But the matrix  $L=D-GH$ is not a Hermitian matrix in
general.This is because the matrix  $G$ is a real symmetric
matrix while the matrix $H$ is a Hermitian matrix, and we know
that the product of the real symmetric matrix and the hermitian
matrix is not going to give a resultant hermitian matrix all the
times. If we observe the matrix $G$ we see that the (i,j) th
element of the matrix is given by the inner product $\langle
\psi_{i}(\theta_i)|\psi_{j}(\theta_j)\rangle=
\cos(\frac{\theta_i}{2})\cos(\frac{\theta_j}{2})+
\sin(\frac{\theta_i}{2})\sin(\frac{\theta_j}{2})$, which is a
real quantity. Here we clearly see that in the matrix $G$ (i,j)th
element is equal to (j,i) th element , with one as the principal
diagonal entries
.Hence the matrix  $G$ is a real symmetric matrix \\
In the matrix $H$ the (i,j) th entry is given by, $\langle
\Sigma(\phi_i)|\Sigma(\phi_j)\rangle=\frac{1}{2}[1+e^{\imath(\phi_j
-\phi_i)}]$ which is a complex quantity and here we see that the
(j,i) th entry of the matrix $H$ is conjugate of the (i,j) th
entry, with one as principal diagonal entries. From here we
conclude that the matrix $H$ is a Hermitian matrix.\\

In general, the principal diagonal elements of the matrix $L=D-GH$ is given by,\\
$L_{ii}=\langle
\psi_i(\theta_i,\phi_i)|\psi_i(\theta_i,\phi_i)\rangle-\gamma_i(\langle\psi_i(\theta_i)|\psi_i(\theta_i)\rangle+\sum_{j=1,j\neq
i}^{n}
\langle \psi_i(\theta_i)|\psi_j(\theta_j)\rangle \langle \Sigma_j(\phi_j)|\Sigma_i(\phi_i)\rangle)$.\\
The off diagonal elements of the matrix $L$ is given by,\\
$L_{ij}= \langle
\psi_i(\theta_i,\phi_i)|\psi_j(\theta_j,\phi_j)\rangle-\sqrt{\gamma_i\gamma_j}(\sum_{k=1}^{n}\langle
\psi_i(\theta_i)|\psi_k(\theta_k)\rangle \langle
\Sigma_k(\phi_k)|\Sigma_j(\phi_j)\rangle)$.\\ Unless the matrix
$L$ is hermitian , we cannot have the corresponding quadratic
form. For the matrix $L$ to be Hermitian , we must have
\\
(i) $ L_{ij}=L_{ji}^{*}$, (where $L_{ji}^{*}$ is the conjugate of
$L_{ji}$)\\
(ii) $L_{ii}$ will be a real quantity. \\In general the matrix L
is not Hermitian as the elements $L_{ii}$ are in general complex
quantities, as a consequence of which we don't have the
corresponding quadratic form, and henceforth there arise no
question for showing $L$ as a positive definite matrix. However
under certain conditions we can show the matrix $L$ to be
positive definite.\\
To show the matrix $L$ to be hermitian we must have to show only
the principal diagonal elements $L_{ii}$ are real, as
$L_{ij}=L_{ji}^{*}$. Now the diagonal elements $L_{ii}$ are real
iff $\sum_{j=1,j\neq i}^{n} \langle
\psi_i(\theta_i)|\psi_j(\theta_j)\rangle Im[\langle
\Sigma_j(\phi_j))|\Sigma_i(\phi_i)\rangle])=0$ and hence the
principal diagonal elements reduces to $L_{ii}=\langle
\psi_i(\theta_i,\phi_i)|\psi_j(\theta_j,\phi_j)\rangle-\gamma_i(\langle
\psi_i(\theta_i)|\psi_j(\theta_j)\rangle+\sum_{j=1,j\neq i}^{n}
\langle\psi_i(\theta_i)|\psi_j(\theta_j)\rangle \\Re[\langle
\Sigma_j(\phi_j)|\Sigma_i(\phi_i)\rangle])$. Now if we consider
the corresponding quadratic form of the matrix $L$,
\begin{eqnarray}
XLX^{t}=(x_1, x_2,....,x_n )[L_{ij}](x_1, x_2,....,x_n )^{t}=
\sum_{i=1}^n L_{ii}x_i^{2}+\sum_{i}\sum_{j}L_{ij}x_{i}x_{j}
\end{eqnarray}
Now the matrix L is positive definite only when the above
expression (10) is positive. This is possible only under the
following conditions:\\
(i)$L_{ii}=\langle
\psi_i(\theta_i,\phi_i)|\psi_i(\theta_i,\phi_i)\rangle-\gamma_i(\langle\psi_i(\theta_i)|\psi_i(\theta_i)\rangle+\sum_{j=1,j\neq
i}^{n}
\langle \psi_i(\theta_i)|\psi_j(\theta_j)\rangle \langle \Sigma_j(\phi_j)|\Sigma_i(\phi_i)\rangle))> 0$\\
(ii) $\sum_{i=1}^n
L_{ii}x_i^{2}>\sum_{i}\sum_{j}L_{ij}x_{i}x_{j}\Longrightarrow
\sum_i[1-\gamma_i-\gamma_i\sum_{j=1,j\neq i}^{n} \langle
\psi_i(\theta_i)|\psi_j(\theta_j)\rangle \langle
\Sigma_j(\phi_j)|\Sigma_i(\phi_i)\rangle]x_i^{2}
>\sum_i\sum_j[\langle
\psi_i(\theta_i,\phi_i)|\psi_j(\theta_j,\phi_j)\rangle-\sqrt{\gamma_i\gamma_j}(\sum_{k=1}^n
\langle \psi_i(\theta_i)|\psi_k(\theta_k)\rangle\langle
\Sigma_k(\phi_k)|\Sigma_j(\phi_j)\rangle)]x_ix_j$~~$\forall
x_i,x_j$. If the above conditions are satisfied, then the matrix
$L$ will be a positive definite matrix. As a consequence of which
we can say that the equation (7) will be satisfied by positive
definite matrix $\Gamma$. Hence under these conditions the
converse of the \textit{Theorem1} will be satisfied, henceforth
unitary evolution (2) will hold and equation (1) can be realized
in physics. This clearly indicates that there are certain class
of states on Bloch sphere satisfying the above conditions for
which the information splitting will be possible.

\section{\bf Conclusion}
In summary we can say that there is no possibility of splitting
the quantum information either deterministically or
probabilistically. The result obtained here is interesting in the
sense that it will help us to understand and to classify the
impossible operations in quantum information theory more
specifically. As a consequence of which one can make a comment
that splitting  of quantum information are different from cloning
and deletion in the sense that these operations unlike cloning
and deletion cannot be achieved even probabilistically. However
this doesn't rule out the probabilistic quantum information
splitting of certain class of states under certain restricted
conditions. This also doesn't rule out the possibility of
approximate splitting of quantum information  .
\section{\bf Acknowledgement}
I.C acknowledge Prof C.G.Chakraborti for being the source of
inspiration in carrying out research. Authors acknowledge
S.Adhikari for having various useful discussions.
\section{\bf Reference}
$[1]$ W.K.Wootters and W.H.Zurek,Nature \textbf{299},802(1982).\\
$[2]$ H.P.Yuen, Phys.Lett.A \textbf{113}, 405(1986).\\
$[3]$ V.Buzek and M.Hillery, Phys.Rev.A \textbf{54}, 1844(1996) \\
$[4]$ N.Gisin and B.Huttner, Phys. Lett. A \textbf{228}, 13(1997) \\
$[5]$ S.Adhikari, A.K.Pati, I.Chakrabarty, B.S.Choudhury,
\textit{Hybrid
Cloning Machine} (under preparation).\\
$[6]$ L. M.Duan and G.C.Guo, Phys. Rev. Lett. \textbf{80}, 4999(1998)\\
$[7]$ A.K.Pati and S.L.Braunstein, Nature \textbf{404},164(2000)\\
$[8]$ A.K.Pati and S.L.Braunstein, e-print quant-ph/0007121\\
$[9]$ Jian Feng. et.al, Phys. Rev. A \textbf{65}, 052311(2002)\\
$[10]$ A.K.Pati, Phys.Rev.A \textbf{66},062319 (2002)\\
$[11]$ Wei Song et.al, Physics Letters A, \textbf{330}, 155-160
(2004)
\\
$[12]$ Duanlu Zhou, Bei Zeng, and L. You, e-print quant-ph/0503168\\
$[13]$ A.K.Pati and Barry C.Sanders, e-print quant-ph/0503138\\
$[14]$ I.Chakrabarty, S.Adhikari, Prashant, B.S.Choudhury ,
\textit{Inseparability of Quantum Parameters} (communicated)\\
$[15]$ D.Qiu, Phys.Lett.A  \textbf{301},112 (2002)\\
$[16]$ S. Adhikari, Phys. Rev. A \textbf{72}, 052321 (2005\\
\end{document}